\RequirePackage{fix-cm}

\documentclass{svjour3}                     
\smartqed  

\usepackage{graphicx}
\usepackage{url}

\usepackage{epsfig}
\usepackage{amssymb,amsmath}
\usepackage{color}
\usepackage{textcomp}
\usepackage{microtype}
\usepackage{float}
\usepackage{subfloat}
\usepackage{epstopdf}
\usepackage{longtable}
\usepackage{rotating}
\usepackage{array}
\usepackage{multicol}
\usepackage{multirow}

\begin{document}

\title{Mobile Multi-View Object Image Search}

\author{\mbox{Fatih \c{C}al{\i}\c{s}{\i}r \and Muhammet Ba\c{s}tan \and \"{O}zg\"{u}r Ulusoy \and U\u{g}ur G\"{u}d\"{u}kbay}}

\authorrunning{\c{C}al{\i}\c{s}{\i}r \ et al.} 

\institute{F. \c{C}al{\i}\c{s}{\i}r, \"{O}. Ulusoy, U. G\"{u}d\"{u}kbay \at
	Bilkent University, Department of Computer Engineering, Bilkent 06800 Ankara, Turkey \\	
	\email{fatih.calisir@bilkent.edu.tr, oulusoy@cs.bilkent.edu.tr, gudukbay@cs.bilkent.edu.tr}  
	\and
	M. Ba\c{s}tan \at
	Nanyang Technological University, School of Electrical and Electronic Engineering, Singapore \\
	\email{mubastan@gmail.com}}

\date{Date: 2017}

\maketitle

\begin{abstract}

High user interaction capability of mobile devices can help improve the accuracy of mobile visual search systems. At query time, it is possible to capture multiple views of an object from different viewing angles and at different scales with the mobile device camera to obtain richer information about the object compared to a single view and hence return more accurate results.
Motivated by this, we propose a new multi-view visual query model on multi-view object image databases for mobile visual search.
Multi-view images of objects acquired by the mobile clients are processed and local features are sent to a server, which combines the query image representations with early/late fusion methods and returns the query results. We performed a comprehensive analysis of early and late fusion approaches using various similarity functions, on an existing single view and a new multi-view object image database.
The experimental results show that multi-view search provides significantly better retrieval accuracy compared to traditional single view search.

\end{abstract}

\section{Introduction}
\label{intro}

Smart mobile devices have become ubiquitous. They are changing the way people access information.
They have some advantages and disadvantages, compared to regular PCs. The advantages are higher accessibility, easier user interaction and the ability to provide context information (e.g., location) using extra sensors, like GPS and compass. The disadvantages are limited computational power, storage, battery life and network bandwidth~\cite{39}, although these are constantly being improved and will be less of an issue in the future.

One traditional way to access information on a mobile device is via text search, by entering a few keywords as query (query-by-keyword); but this is usually cumbersome and slow, considering the small screen size of mobile devices.
As a convenience, it is also possible to initiate text queries via speech, if automatic speech recognition is available. Sometimes, it is very difficult to express a query using only keywords. For instance, when a user at a shoe store wants to know more about a specific type of shoe (e.g., cheaper prices elsewhere, customer reviews), she cannot easily formulate a text query to express her intent. It is much easier to take a photo of the shoe with the mobile device camera, initiate a visual search and retrieve visually similar results. This is now possible, owing to the recent hardware/software developments in mobile device technology, which turned the smart phones with high-resolution cameras, image processing capabilities and Internet connection into indispensable personal assistants. This in turn triggered research interest in mobile visual search and recognition~\cite{7,14,15}, and motivated the industry to develop mobile visual search applications, such as Google Goggles~\cite{16}, CamFind~\cite{4}, Nokia Point \& Find~\cite{31}, Amazon Flow~\cite{AmazonFlow}, Kooaba image recognition~\cite{Kooaba}.

The main focus of this work is to leverage the user interaction potential of mobile devices to achieve higher visual search performance, and hence, provide the users with easier access to richer information.
One potential application area is \textit{mobile product search}.
When the user wants to search for a specific object, she can take a photo of the object to initiate a visual search. Additionally, she can easily tap on the screen to mark the object-of-interest and provide extra information to the search system to suppress the influence of background in matching~\cite{tapTell}.
More importantly, the user can take multiple photos of the object-of-interest from different viewing angles and/or at different scales, thereby providing much more information about the query object. We refer to \textit{multi-view object image search} as providing multiple query images of an object from various viewing angles and at various scales and combining the query images using early/late fusion strategies to achieve higher retrieval precision on single-- and/or multi-view object image databases.
High precision on mobile retrieval is especially critical because the screen is too small to display many results, and more importantly, the user usually does not have much time and patience to check more than $10-20$ results.

\begin{figure}[h!]
	\centering
	\includegraphics[width=0.60\textwidth]{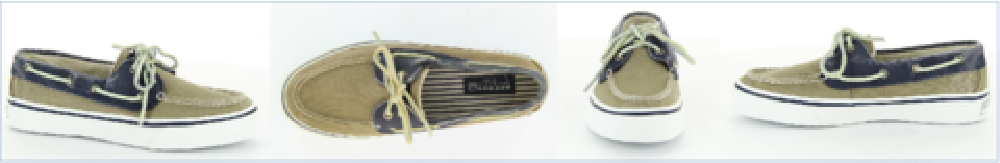}\\
	\includegraphics[width=0.60\textwidth]{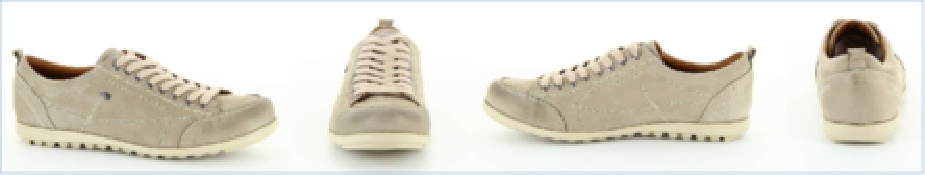}
	\caption{Multi-view images of two different shoes. Online stores typically contain multi-view images of products.}
	\label{fig:MultiViewObjectImages}
\end{figure}

\begin{figure}[h!]
	\centering
	\includegraphics[width=0.9\textwidth]{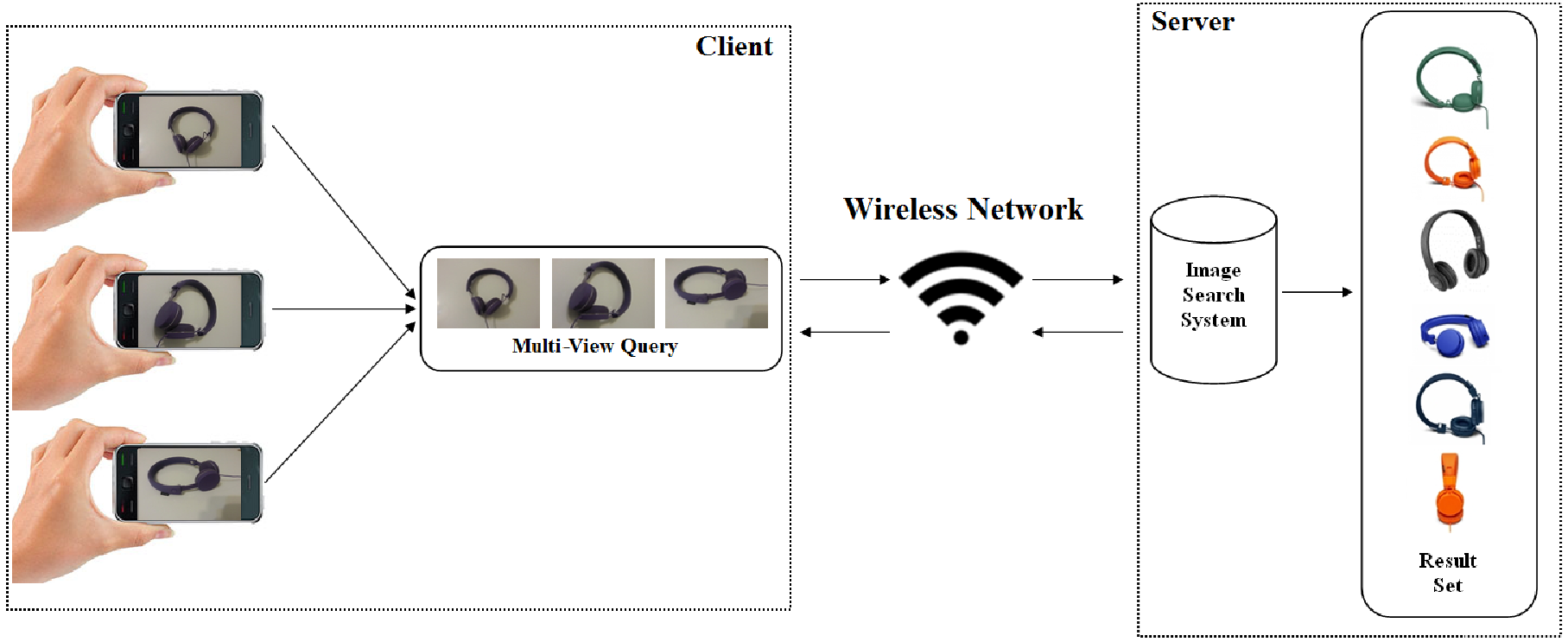}
	\caption{Client-server architecture of our mobile multi-view visual search system.}
	\label{fig:System-Architecture}
\end{figure}

Multi-view search is different from multi-image search. In multi-image search, multiple images of an object category are used to perform a search~\cite{1,41,49}; the images do not belong to the same instance of the queried object.
In multi-view search, query images belong to the same object instance.
To illustrate the benefits of multi-view object image search, consider the multi-view images of two different shoes in Figure~\ref{fig:MultiViewObjectImages}, taken from four different viewing angles at the same scale. Such images are typical on online stores, i.e., multi-view images of objects on clean backgrounds. Assuming the database contains such multi-view images for each object, when a user performs a search using a photo that is close to one of the available views, the results she will get will be better than when the query image has a completely different view. Intuitively, if the user takes multiple photos of the object from different viewing angles, the chance that the query images are similar to the ones in the database will increase. This is also valid when the database contains single view images of each object. The effect of multiple scales is similar. In summary, at query time, the user does not know the view and scale of the object images in the database; by taking multiple photos from different views and at different scales, she can increase the chance of capturing views similar to the database images. This is enabled by the interactivity of the mobile devices; on a non-mobile platform, like a PC, it would be difficult to obtain such multi-view images of an object and perform a multi-view query. Therefore, such a multi-view search system is most suitable for mobile devices with a camera.

In this paper, we address the following questions concerning multi-view object image search:

\begin{itemize}

 \item Is a multi-view object image database better than a single view database in terms of retrieval precision, and if so how much?

 \item Do multi-view queries improve retrieval precision on single view or multi-view object image databases, and if so how much?

 \item Are multi-view queries better than multi-image queries in terms of retrieval precision, and if so how much?

 \item Multi-view queries need special treatment to combine multiple query/data\-base images using early/late fusion strategies~\cite{EarlyLateFusion,Fusion}. What are the best similarity functions and early/late fusion methods for a search system employing multi-view queries or databases?

 \item What is the additional computational cost of multi-view search, and is the performance improvement worth the additional cost?

\end{itemize}

To the best of our knowledge, there is no work describing a multi-view object image search system, addressing these issues. We show through extensive experiments that multi-view queries and/or databases improve retrieval precision significantly, over both single view and multi-image queries, at a cost of modest increase in computation time due to the increase in the number of images to be processed.

To demonstrate the benefits of multi-view search, we built a mobile visual search system based on client-server architecture (cf. Figure~\ref{fig:System-Architecture}), using the well-known bag-of-visual-words (BoW) approach~\cite{14,15}. We constructed a multi-view object image database and performed extensive experiments on both single and multi-view object image databases with single, multi-image and multi-view queries using various similarity functions and fusion strategies, and  presented the results in a systematic way.

\section{Related Work}
\label{related}

Due to the recent advances in mobile devices with cameras, there has been a growing interest in mobile visual search. Research works investigate different aspects of mobile visual search, such as architectures, power efficiency, speed, and user interaction. Chen and Girod~\cite{7} describe a mobile product recognition system where the products are CDs, DVDs and books that have printed labels. The system is local feature based, and Compressed Histogram of Gradients (\textit{CHOG}) and Scale-Invariant Feature Transform (\textit{SIFT}) are used as local features. Two client-server architectures are implemented and compared in terms of response time: one is sending images, the other one is extracting features on the client and sending the features. Sending features took five seconds, sending images took ten seconds to respond. This means that over slow connections like 3G it is faster to extract and send features. Based on this finding, we preferred the former approach in our implementation (Figure~\ref{fig:MobileSystem}).

Storage space and retrieval speed are critical in mobile visual search. Girod et~al.~\cite{15} describe a mobile visual search system that adopts the client-server architecture in which the database is stored on the phone. The system uses the BoW approach, and four different compact database methods are experimented and their performances are compared. Li et al.~\cite{emod} propose an on-device mobile visual search system. The system uses the BoW approach with a small visual dictionary due to the memory limitation. Additionally, the most useful visual words are selected to decrease the retrieval time considering the processor limitation. Guan et al.~\cite{efficientBOF} describe an on-device mobile image search system, which is based on bag-of-features (BoF) approach. The system uses approximate nearest neighbor search to use high dimensional BoF descriptors on the mobile device with less memory usage. The search system also utilizes the GPS data from the mobile device to reduce the number of images to be compared. In our case, considering the potential application areas of our system (e.g., mobile product search), the database must be stored on the server side. For speeding up the query processing, feature extraction and query processing is run in parallel as the user is taking multiple photos of the query object (see Section~\ref{sec:server-qp}).

Mobile devices have high user interaction potential; this has been utilized for better retrieval.
Joint Search with Image Speech and Words (\textit{JIGSAW})~\cite{45} is a mobile visual search system that provides multimodal queries. This system allows the user to speak a sentence and performs text-based image search. The user selects one or more images from the result set to construct a visual query for content-based image search. In~\cite{34}, a mobile product image search system that automatically extracts the object in the query image is proposed. From the top n images that have a clean background, object masks are found. The object in the query image is then extracted by using a weighted mask approach and its background is cleaned. The cleaned query image is finally used to perform image search. Extracting the object-of-interest and performing the query with a clean background is shown to work better. Similarly, \textit{TapTell}~\cite{tapTell} is an interactive mobile visual search system, in which users take a photo and indicate an object-of-interest within the photo using various drawing patterns. In our system, user interaction is used to obtain multi-view images of the query object. Further user interaction, e.g., to select the object of interest to suppress the effects of background, would further improve the performance as indicated by our experiments (see Section~\ref{sec:mvod-result}).

Landmark and location recognition and search are among popular mobile application areas~\cite{chen2011city,aqs-amm11,aqs-tomccap12,efficientBOF,min2014mobile,zhu2015landmark}. In~\cite{min2014mobile}, 3D models of landmarks are constructed offline; then, low resolution query images taken by the mobile device are sent to the server and matched with 3D landmark models. With 3D models, it is possible to match query images taken from different viewpoints. However it is not easy to construct the 3D models of landmarks, especially large ones, because many views may be needed for a high quality reconstruction. In this work, we use the multi-view images directly, instead of building 3D models, since for a typical mobile product search system, multi-view images of products are readily available (but not as many views as would be needed to reconstruct a 3D model of the product). In~\cite{aqs-amm11,aqs-tomccap12}, an \textit{Active Query Sensing} system is proposed to help the user take a better query image when the first query fails, for mobile location search. The system learns the saliency of each view of a location to determine the discriminability, which is later used for suggesting better views based on the first query. Using multi-view queries, as in our system, might improve the accuracy at the first query and reduce the need to refine the search. However, the idea of selecting discriminative views is promising and can be investigated further for multi-view queries on multi-view databases to find out whether it is better than fusion approaches used in this work.

There are several mobile image search and recognition applications available on the mobile market.
\textit{Point\&Find}~\cite{31} allows the users point the camera to the scene or object and get information about it. \textit{Kooaba} is a domain-specific search system whose target domains are books, DVD and game covers~\cite{Kooaba}. \textit{Digimarc Discover}~\cite{8} is similar to \textit{Point\&Find}; the user points the camera to an object and gets information about it. \textit{PlinkArt}~\cite{47} is another domain-specific mobile search system whose target domain is well-known artworks. The user takes a photo of a well-known artwork and gets information about it. One of the latest mobile search application is \textit{CamFind}~\cite{4}, which is a general object search system. When the user takes a photo of a scene, products are identified and similar objects are listed as a result. Another recent mobile search application is \textit{Amazon Flow}~\cite{AmazonFlow}; the user points the camera to the object and receives information about it. These examples indicate the commercial potential of mobile visual search systems.

Multi-image queries have been used to improve image retrieval. Arandjelovic and Zisserman~\cite{1} propose an object retrieval system using multiple image queries. The user enters a textual query and Google image search is performed using this textual query. The top eight images are then retrieved and used as query images. Early and late fusion methods are applied. Tang and Acton~\cite{41} propose a system that extracts different features from different query images. These extracted features are then combined and used as the features of the final query image. The system proposed in~\cite{29} allows users to select different regions of interest in the image. Then each region is treated as separate queries and their results are combined. Zhang et al.~\cite{51} describe a similar system, which also uses regions; however, these regions are extracted automatically and the user selects among them. Xue et al.~\cite{49} propose a system that uses multiple image queries to reduce the distracting features by using a hierarchical vocabulary tree. The system focuses on the parts that are common in all the query images. The multi-query system described in~\cite{24} uses early fusion; each database image is compared with each query image and each query image gets a weight according to the similarity between the query image and the database image.

All these works use multiple query images on single view databases for performance improvement; they do not utilize multi-view queries on multi-view databases. Moreover, a multi-view object image dataset to evaluate multi-view search systems is not publicly available. This paper aims to fill in these gaps.

\section{Proposed Mobile Visual Search System}
\label{sec:mvs}

The proposed mobile multi-view visual search system is based on the well-known BoW approach: the images are represented as a histogram of quantized local features, called the BoW histogram. First, interest points are detected on images; the points are described with local descriptors computed around the points. Then, a vocabulary (dictionary) is constructed with a set of descriptors from the database, typically using the k-means clustering algorithm. Finally, images are represented as a histogram of local interest point descriptors quantized on the vocabulary. When a query image is received by the search system, local features are extracted and BoW histogram is computed. The query histogram is compared with the histograms stored in the database, and the best \textit{k} results are returned to the user (cf. Figure~\ref{fig:System-Architecture}).

Local features are key to the performance of the search system. In a mobile system, they should also be efficiently computable. To this end, we employed two fast local feature detectors: Harris and Hessian ~\cite{SurfPaper,LocalFeatureSurvey}. They detect two types of complementary local structures on images: \textit{corners} and \textit{blobs}. Using complementary interest points are useful for improving the expressive power of features and hence the retrieval precision. The detected points are represented with the SIFT descriptor. The BoW histograms are computed for Harris and Hessian separately, and then they are concatenated to obtain the BoW histogram of an image.

For ranking, the database images need to be compared with the query image(s), based on the BoW histograms.
It is crucial to select the right similarity functions for high retrieval precision and low computational cost. There are various similarity functions that can be used to compare histograms~\cite{20,22,27}. We experimented with the similarity functions given in Table~\ref{table:SimMetric} and presented a comparison in terms of retrieval precision and running time in Section~\ref{sec:experiments}. In the table, $h^q$ and $h^d$ represent the histogram of the query and database images, respectively. In the formulae, $q_i$ and $d_i$ are the $i^{th}$ histogram bin of query and database histograms, respectively.

\begin{table}[h]
	\caption{Similarity functions for comparing BoW histograms.}
	\centering
	\begin{tabular}{l c c}
		\hline\hline
		Similarity Function & Symbol & Formula \\ [0.5ex]		
		\hline
		\\[-1.5ex]
		Dot Product~\cite{22} & \mbox{\textit{dot}}$(h^q,h^d)$ & $\sum_iq_id_i$ \\ [2ex]
		Histogram Intersection~\cite{27} & \mbox{\textit{HI}}$(h^q,h^d)$ & $\dfrac{\sum_imin(q_i,d_i)}{min(|h^q|,|h^d|)}$ \\ [5ex]
		Normalized Histogram\\Intersection~\cite{22} & \mbox{\textit{NHI}}$(h^q,h^d)$ & $\sum_imin\bigg(\dfrac{q_i}{\sum_iq_i},\dfrac{d_i}{\sum_id_i}\bigg)$\\ [5ex]		 
		Normalized Correlation~\cite{27} & \mbox{\textit{NC}}$(h^q,h^d)$ & $\dfrac{\sum_i{q_i}{d_i}}{{\sqrt{{\sum_i}{q_i^2}}}\times{\sqrt{{\sum_i}{d_i^2}}}}$ \\ [5ex]			 
		Min-Max Ratio~\cite{MinMaxRatio} & \mbox{\textit{MinMax}}$(h^q,h^d)$ & $\dfrac{\sum_imin(q_i,d_i)}{\sum_imax(q_i,d_i)}$\\
		[3ex]
		\hline
	\end{tabular}
	\label{table:SimMetric}
\end{table}

\subsection{Multi-View Search}
\label{sec:multivs}

\begin{figure}[h!t]
    \centering
    \includegraphics[width=0.80\textwidth]{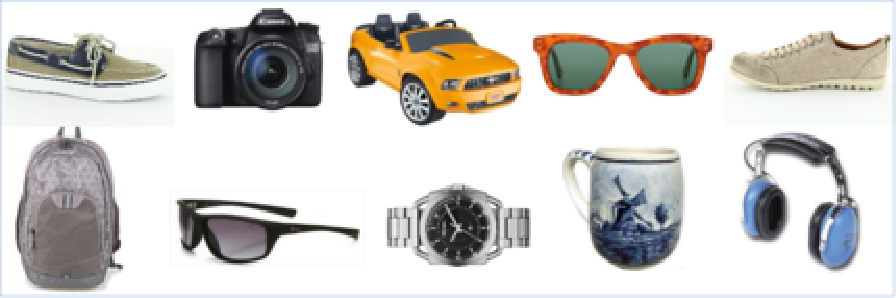}
    \caption{Single view images: each image is a typical, single view of an object.}
    \label{fig:single-view-images}
\end{figure}

\begin{figure}[h!t]
    \centering
    \includegraphics[width=0.80\textwidth]{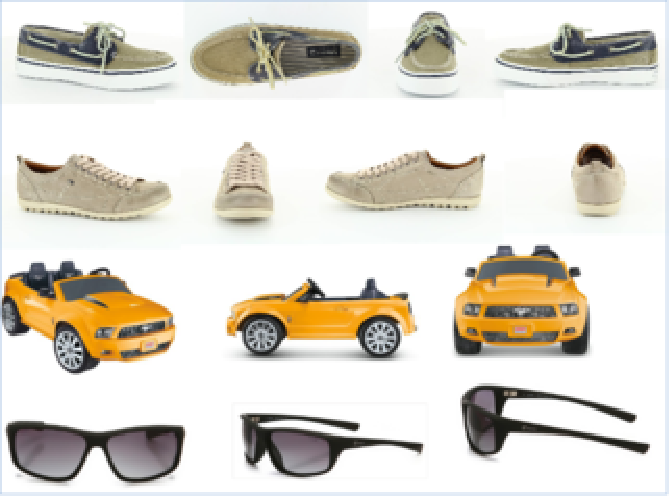}
    \caption{Multi-view images: each object has multiple images from different viewing angles (and/or at different scales).}
    \label{fig:multi-view-images}
\end{figure}

Image databases typically contain single view images of objects or scenes, as in Figure~\ref{fig:single-view-images}. At query time, if the user captures and uses a view close to the one in the database, she will retrieve the relevant image, but the user does not have any idea about the view stored in the database. If the query image has a slightly different view or scale, the invariancy of local features can handle such view/scale changes; but if the view/scale change is significant, the system will most probably fail. As a solution, the user may take multiple photos from different viewing angles and at different scales to increase the chance of providing query images similar to the database images. Moreover, if the database images are also multi-view, we can expect to get even better results. Hence, both the query and database images can be multi-view, each object/scene having multi-view images, as in Figure~\ref{fig:multi-view-images}. In the most general case, the query may contain $M \geq 1$ images of an object and the database may consist of $N \geq 1$ images of each object.

\begin{description}
	\item
	\textit{Single-view query and single-view database ($M=1$, $N=1$)}: Both the query and database objects have a single image that represents a specific view of the object, as in Figure~\ref{fig:single-view-images}. During retrieval, the query image is compared to every database image using a similarity function (cf. Table~\ref{table:SimMetric}) to find best \textit{k} matches.
	
	\item
	\textit{Single-view query and multi-view database ($M=1$, $N \geq 1$)}: The query has single-view (cf. Figure~\ref{fig:single-view-images}) and database objects have multi-view images (cf. Figure~\ref{fig:multi-view-images}). During retrieval, early/late fusion methods (cf. Sections~\ref{sec:earlyfusion} and \ref{sec:latefusion}) are employed to find and return best \textit{k} matching database objects.
	
	\item
	\textit{Multi-view query and single-view database ($M \geq 1$, $N=1$)}: The query has multi-view images, the database has a single image for each object. During retrieval, early/late fusion methods are employed to find and return best \textit{k} matching database images.	
	
	\item
	\textit{Multi-view query and multi-view database ($M \geq 1$, $N \geq 1$)}: Both the query and database objects have multi-view images. This is the most general case and comprises the previous three cases. During retrieval, early/late fusion methods are employed to find and return best \textit{k} matching database objects. We expect to get the best retrieval precision, but at an increased computational cost.
	
\end{description}

When the query or database objects have multiple images, we must employ fusion methods to process the queries and find best \textit{k} matching database objects. This is one of the crucial steps to achieve high retrieval performance. There are mainly two types of fusion methods: \textit{early fusion} and \textit{late fusion}. We performed comprehensive experimental analysis of several early and late fusion methods.

\subsubsection{Early Fusion}
\label{sec:earlyfusion}

Early fusion, also referred to as fusion in feature space, is the approach in which the BoW histograms of multiple images are combined into a single histogram and the similarity function is applied on the combined histograms.
We used the early fusion methods given in Table~\ref{table:HistogramMethods}~\cite{1}. In the table, the histograms for $M$ images are combined into ${h^c}$; ${h^j_i}$ is the $i^{th}$ bin of histogram $h^j$ of image $j$.

\begin{table}[h]
	\caption{Early fusion methods.}
	\centering
	\begin{tabular}{c c l}
		\hline\hline
		Method &  Formula \\ [0.5ex]
		\hline
		\rule{0pt}{5ex}
		Sum Histogram & $h^c_i = \sum_{j=1}^M h^j_i$ \\ [2ex]
		Average Histogram &  $h^c_i = \dfrac{\sum_{j=1}^M h^j_i}{M}$ \\ [2ex]
		Maximum Histogram & $h^c_i = max(h^1_i, \ldots, h^{M}_i)$ \\ [2ex]		
		\hline
	\end{tabular}
	\label{table:HistogramMethods}
\end{table}

\subsubsection{Late Fusion}
\label{sec:latefusion}

Late fusion, also referred to as decision level fusion,
considers each query and database image separately to obtain similarity scores between the query and database images using their BoW histograms; the final result list is obtained by combining the individual similarity scores. This can be done in two ways: (1) \textit{image similarity and ranking} and (2) \textit{image set similarity and ranking}.

\noindent
\textit{\textit{Image similarity and ranking.}} The image histograms in the query are compared to all the image histograms of all the objects in the database; a single result list is obtained by ranking the database objects based on similarity scores or ranking. We used the following methods~\cite{24,52}.

\begin{itemize}
 \item \textit{Max Similarity (\textit{MAX SIM}).} Each database image is compared with the query images and the similarity is taken as the maximum of the similarities.
 \item \textit{Weighted Similarity.} Each database image is ranked according to a weighted similarity to the query images.
 \item \textit{Count.} For multiple query images, multiple result lists are obtained. Then, for each image, a counter is incremented if it is in a list. Finally, the counter value is used to rank the database images (higher value, higher rank).
 \item \textit{Highest Rank.} For multiple query images, multiple result lists are obtained and the highest rank is taken for each database image.
 \item \textit{Rank Sum.} For multiple query images, multiple result lists are obtained and the ranking of each image in every list is summed and the resulting values are used to rank the database images.
\end{itemize}

\begin{figure}[h!]
	\centering
	\includegraphics[width=1\textwidth]{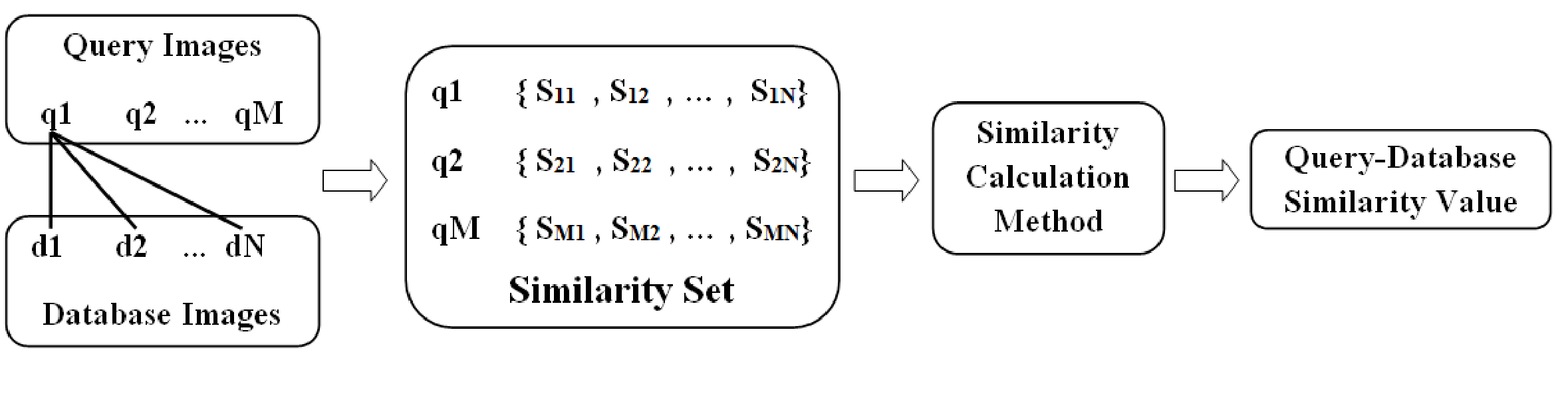}
	\caption{Similarity computation between image sets. The query has $M$ images, the database object has $N$ images. A similarity score $S_{ij}$ is computed between every query image \textit{i} and every database object image \textit{j}, resulting in $M \times N$ similarity scores.}
	\label{fig:image-set-similarity}
\end{figure}

\noindent
\textit{\textit{Image set similarity and ranking.}} First, the similarity scores between $M$ images of the query object and $N$ database images of each object are computed, resulting in $M \times N$ similarity scores, as shown in Figure~\ref{fig:image-set-similarity}. Then, an \textit{image set similarity score} between the query object and each database object is computed, and finally, database objects are ranked according to the image set similarity scores.

The image set similarity scores between  $M$ query images and $N$ database object images are computed in one of the following ways, based on the individual similarity scores between the query and database images (cf. Figure~\ref{fig:image-set-similarity}).

\begin{itemize}

 \item \textit{Maximum Similarity (MAX).} The similarity score is the maximum of all $M \times N$ similarity scores.

    \begin{equation*}
      \displaystyle
	Similarity = max(S_{ij})
    \end{equation*}

  If at least one of the query images is very similar to one of the database object images, this measure will return good results.

 \item \textit{Average Similarity (AVERAGE).} The similarity score is computed as the average of all $M \times N$ similarity scores.
      \begin{equation*}
	\displaystyle
	  Similarity = \dfrac{{\displaystyle\sum^M_{i=1}}~{\displaystyle\sum^N_{j=1}}~{S_{ij}}}{M \times N}
      \end{equation*}
 The average operator reduces the effects of outliers, but it also reduces the effects of good matches with high similarity scores.

 \item \textit{Weighted Average Similarity (WEIGHTED AVERAGE).} To promote the influence of good matches with high similarity scores, a weight is assigned to each score.

	    \begin{equation*}
		\displaystyle
		\begin{split}		
		    W_{ij} &= \dfrac{S_{ij}}{ {\displaystyle\sum^M_{i=1}}~{\displaystyle\sum^N_{j=1}}~{S_{ij}} }\\		
		    Similarity &= {\sum^M_i}~{\sum^N_j}~{S_{ij}}\times{W_{ij}}
		\end{split}
	    \end{equation*}

 \item \textit{Average of Maximum Similarities (AVERAGE MAX).} First, the maximum similarity for each of $M$ query images to $N$ database object image is computed. Then, the average of $M$ maximum similarity values is computed as the image set similarity.

	    \begin{equation*}
	      \displaystyle
	      Similarity = \dfrac{ \sum^M_i {\max (S_{i1}, \ldots,S_{iN} ) } } {M}
	    \end{equation*}

 \item \textit{Weighted Average of Maximum Similarities (WEIGHTED AVERAGE MAX).} This is similar to the previous method; this time, the average is weighted.

      \begin{equation*}
	    \displaystyle
	    \begin{split}
	    S_i    &= \max (S_{i1}, \ldots,S_{iN} )\\
	    W_{i} &= \dfrac{S_{i}}{ {\sum^M_i}{S_{i}} }\\
	    \displaystyle
	    Similarity &= {\sum^M_i}{W_i}\times{S_i}
	    \end{split}
      \end{equation*}

\end{itemize}

\subsection{Speeding up Multi-View Query Processing}
\label{sec:server-qp}

\begin{figure}[h!]\centering
	\includegraphics[width=1\textwidth]{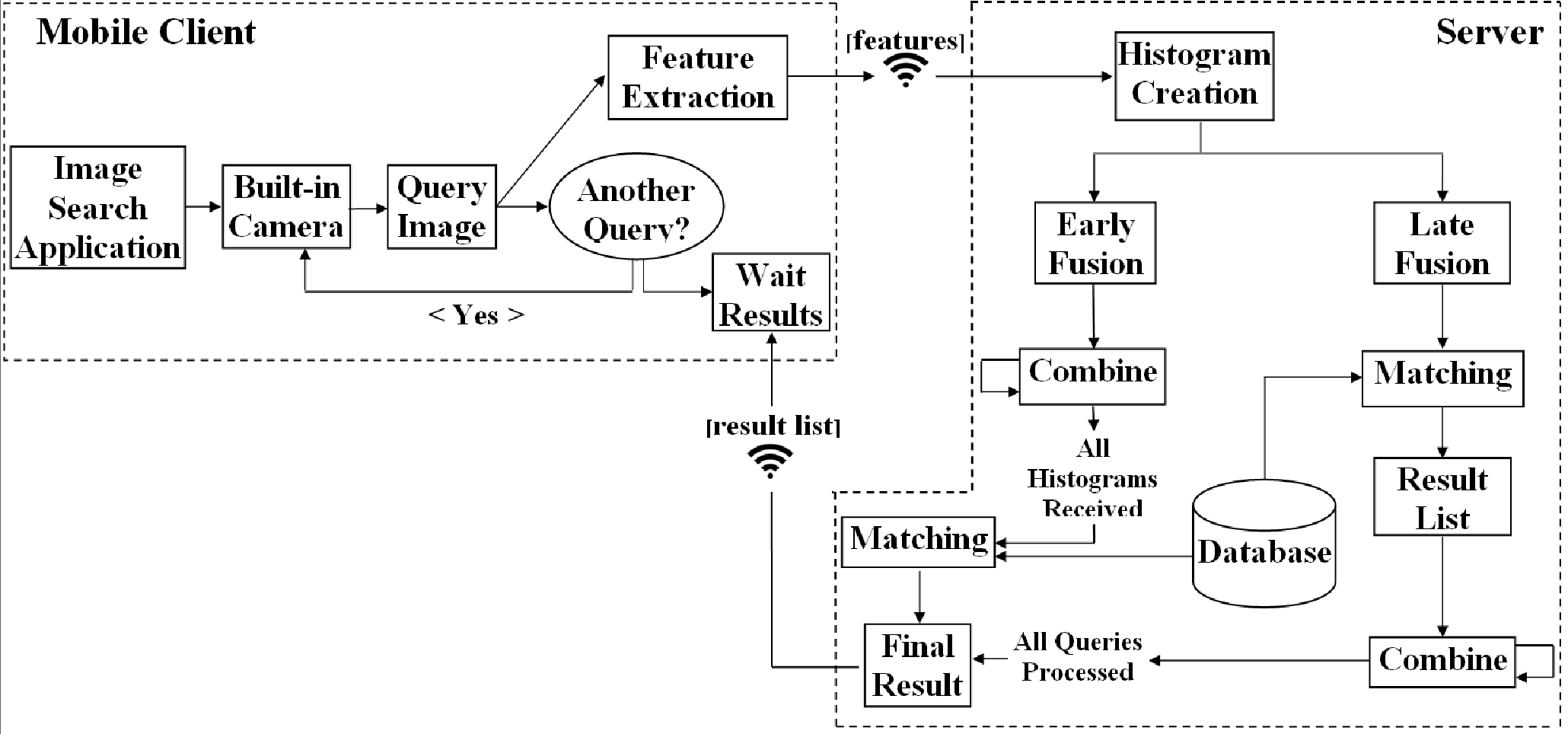}
	\caption{Workflow of our image search system using early and late fusion methods.}
	\label{fig:MobileSystem}
\end{figure}

Multi-view queries are inherently computationally more expensive than single view queries.
However, it is possible to speed up the multi-view search in a mobile multi-view search setting.
As the user is taking multiple photos of the query object, the feature extraction and query processing can run in parallel in the background. This is possible because current mobile devices usually have multi-code processors. While one thread handles photo-taking, another thread can extract and send features to the server, which can start query processing as soon as it receives the features for the first query image. Figure~\ref{fig:MobileSystem} shows the flow diagram of the whole process as implemented in our mobile search system.

\section{Datasets}
\label{sec:data}

We used two different datasets to evaluate the performance of our mobile search system: (i) an existing single view mobile product image search dataset, \textit{Caltech-256 Mobile Product Search Dataset}~\cite{34}, and (ii) a new multi-view object image dataset we constructed for this work.

\begin{figure}[h!]\centering
	\fbox{\includegraphics[width=0.6\textwidth]{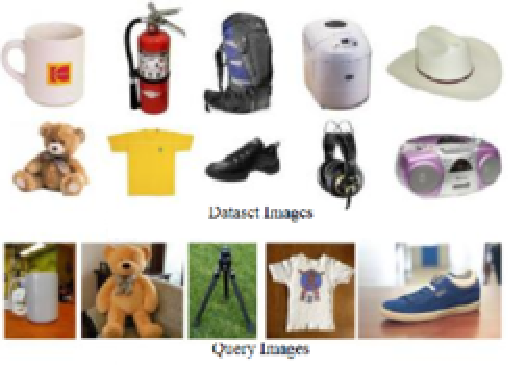}}
	\caption{Sample images from the \textit{Caltech-256} mobile product search dataset.}
	\label{fig:Caltech256}
\end{figure}

\noindent
\textit{(i) \textit{Caltech-256 Mobile Product Search Dataset.}} This is a subset of the Caltech-256 Object Category Dataset~\cite{Caltech256}, which is used to evaluate the performance of the mobile product search system described in~\cite{34}. The dataset has $20$ categories and $844$ object images with clean background; objects are positioned at the image center. There are $60$ query images from six categories; query images contain background clutter. The original Caltech-256 dataset images were downloaded from Google Images. Figure~\ref{fig:Caltech256} shows sample images from the dataset. This is a single view object image dataset. Although the dataset contains multiple images of each category, the images are not multiple views of the same object, rather they are from different objects of the same category.

\begin{figure}[h!]\centering
	\fbox{\includegraphics[width=0.6\textwidth]{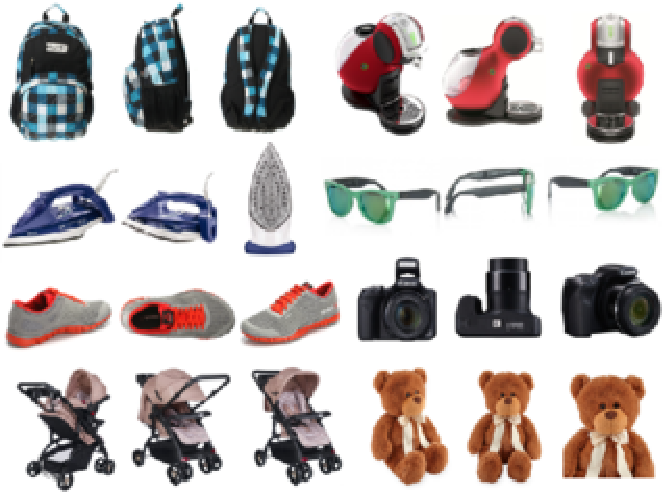}}
	\fbox{\includegraphics[width=0.6\textwidth]{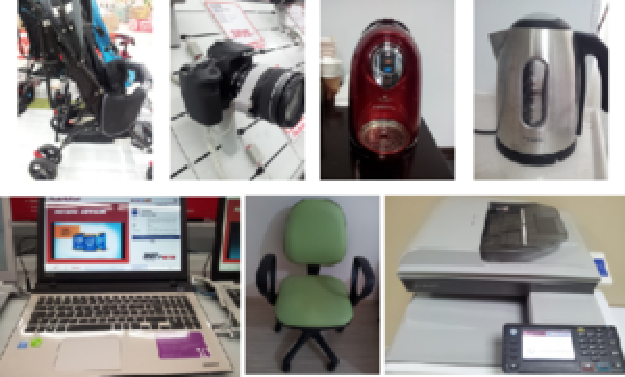}}
	\caption{Sample images from the \textit{MVOD} dataset. Top: database, bottom: queries (selected single view images from multi-view queries).}
	\label{fig:mvod-dataset}
\end{figure}

\noindent
\textit{(ii) Multi-View Object Images Dataset (MVOD).} The main focus of this work is mobile multi-view object image search; hence it is crucial to have a suitable multi-view object image dataset to evaluate the performance of our system. To the best of our knowledge, such a dataset is not publicly available.
We constructed a new dataset, called Multi-View Object Image Dataset (\textit{MVOD 5K}), from online shopping sites.
The dataset has $5000$ images from $45$ different product categories (shoes, backpacks, eyeglasses, cameras, printers, guitars, pianos, coffee machines, vacuum cleaners, irons, etc.).
There are 1827 different object instances (from 45 categories) and each object has at least two different images taken from different views.
On the average, there are $40$ object instances and $111$ images per category and $3$ views per object.
The images mostly have a clean background and objects are positioned at the image centers; this is on purpose, because the goal is to provide a good image of the product to attract the customers.
The dataset is suitable for a mobile product search system, containing images of daily life items sold on online stores, and hence, it is easy to generate multi-view query images with a mobile device. Figure~\ref{fig:mvod-dataset} shows sample images from the dataset. The dataset and more detailed description are available at \url{www.cs.bilkent.edu.tr/~bilmdg/mvod/}.

\section{Experiments}
\label{sec:experiments}

We performed extensive experiments on the \textit{Caltech-256} and \textit{MVOD} datasets and evaluated the performance of various similarity functions and fusion methods. We used the OpenCV library~\cite{OpenCV} to extract the local features (Harris, Hessian detector with SIFT descriptor).

The evaluation is done based on \textit{average precision (AveP)}~\cite{PrecisionPaper}, as shown below. In the equation, \textit{k} represents the rank in the list of retrieved images and \textit{N} is the length of the list. A retrieved object image is relevant if it belongs to the same object category.

  \begin{equation*}
  \begin{split}
  \label{eq:AvgPrec}
    P(k) = \dfrac{\mbox{\it relevant images}{\;\cap\;}\mbox{\it first\,k\,images}}{k}\\
    rel(k)= \begin{cases}
    1, &\mbox{if } \mbox{\it image k\,is\,relevant} \\
    0, & \mbox{otherwise}
    \end{cases} \\
    \displaystyle
    AveP = \dfrac{{\sum_{k=1}^{N}}({P(k){\times}rel(k)})}{N}
  \end{split}
  \end{equation*}

\subsection{Results on the \textit{Caltech-256} Dataset}
\label{sec:caltech-results}

As mentioned above, \textit{Caltech-256} mobile product search dataset is a single-view dataset.
On this dataset, we performed experiments with three types of queries:
\begin{itemize}
 \item \textit{Single view queries.} Each query is a single object image; the same query images, as in~\cite{34}, are used (six categories, each having ten images). Queries with clean and cluttered background are performed and evaluated separately. We used clean background queries provided by~\cite{34}. They were obtained by segmenting out the objects from the background.
 \item \textit{Multi-image queries.} Each query consists of multiple object images from the same category, however, the images belong to different objects, they are not multiple views of the same object. There are six queries for six categories, and all ten images are used in each multi-image query.
 \item \textit{Multi-view queries.} Each query consists of multi-view images of an object; the images were taken with a mobile phone and hence not from the \textit{Caltech-256} dataset. There are four multi-view queries for four categories, each having five images.
\end{itemize}

\begin{figure}[h!]
    \centering
    \includegraphics[width=\textwidth]{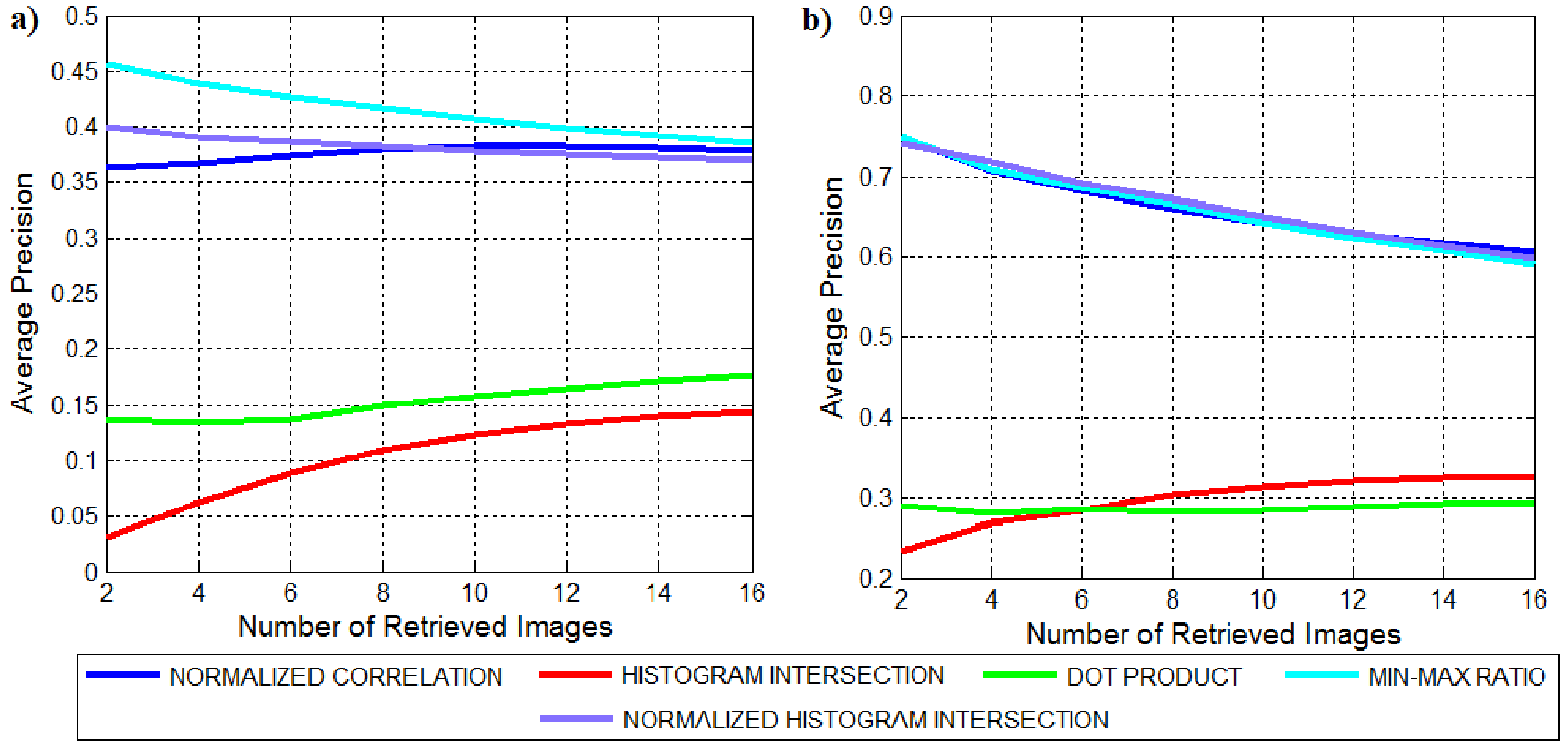}
    \caption{Single view query average precision graphs on \textit{Caltech-256} dataset with various similarity functions: (a) our results with background cluttered queries and (b) our results with clean background queries.}
    \label{fig:AveP-Caltech-single-view}
\end{figure}

\begin{figure}[h!]\centering
	\fbox{\includegraphics[width=1\textwidth]{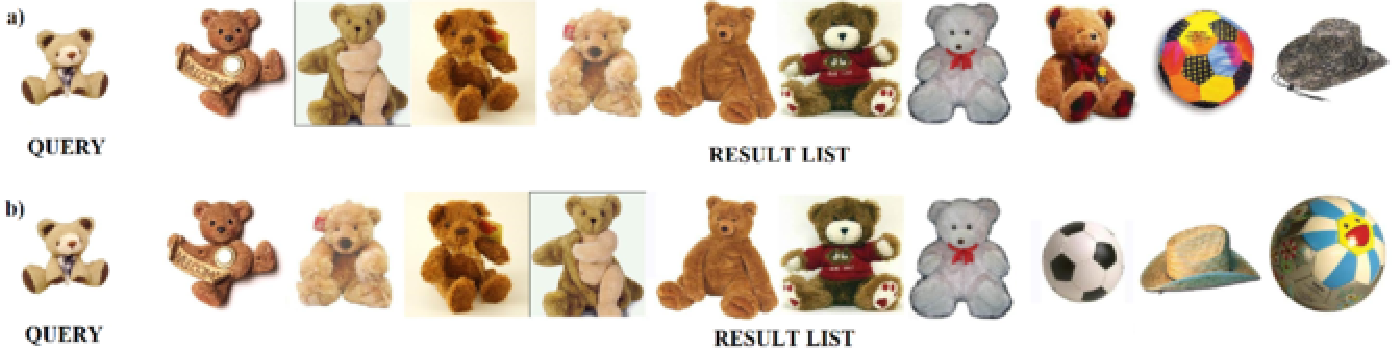}}
	\caption{Single view query examples on \textit{Caltech-256} dataset with two similarity functions: (a) Min-Max Ratio and (b) Normalized Histogram Intersection.}
	\label{fig:Caltech256Fig1Clean}
\end{figure}

The vocabulary size is $3K$ and hard assignment is used for computing the BoW histograms. Figure~\ref{fig:AveP-Caltech-single-view} shows the average precision graphs for single view queries using various similarity functions. The similarity functions \textit{Min-Max Ratio}, \textit{Normalized Histogram Intersection} and \textit{Normalized Correlation} work much better than \textit{Dot Product} and \textit{Histogram Intersection} on both clean and background cluttered queries. As expected, the average precision is higher for queries with a clean background. When we compare our results with those presented in Figure 6~(b) of \cite{34}, our average precision values are $0.1-0.15$ higher than~\cite{34}, probably due to the multiple complementary features (Harris+Hessian with SIFT) we used. Figure~\ref{fig:Caltech256Fig1Clean} shows single view query examples with two different similarity functions.

\begin{figure}[h!]
	\centering
	\includegraphics[width=1\textwidth]{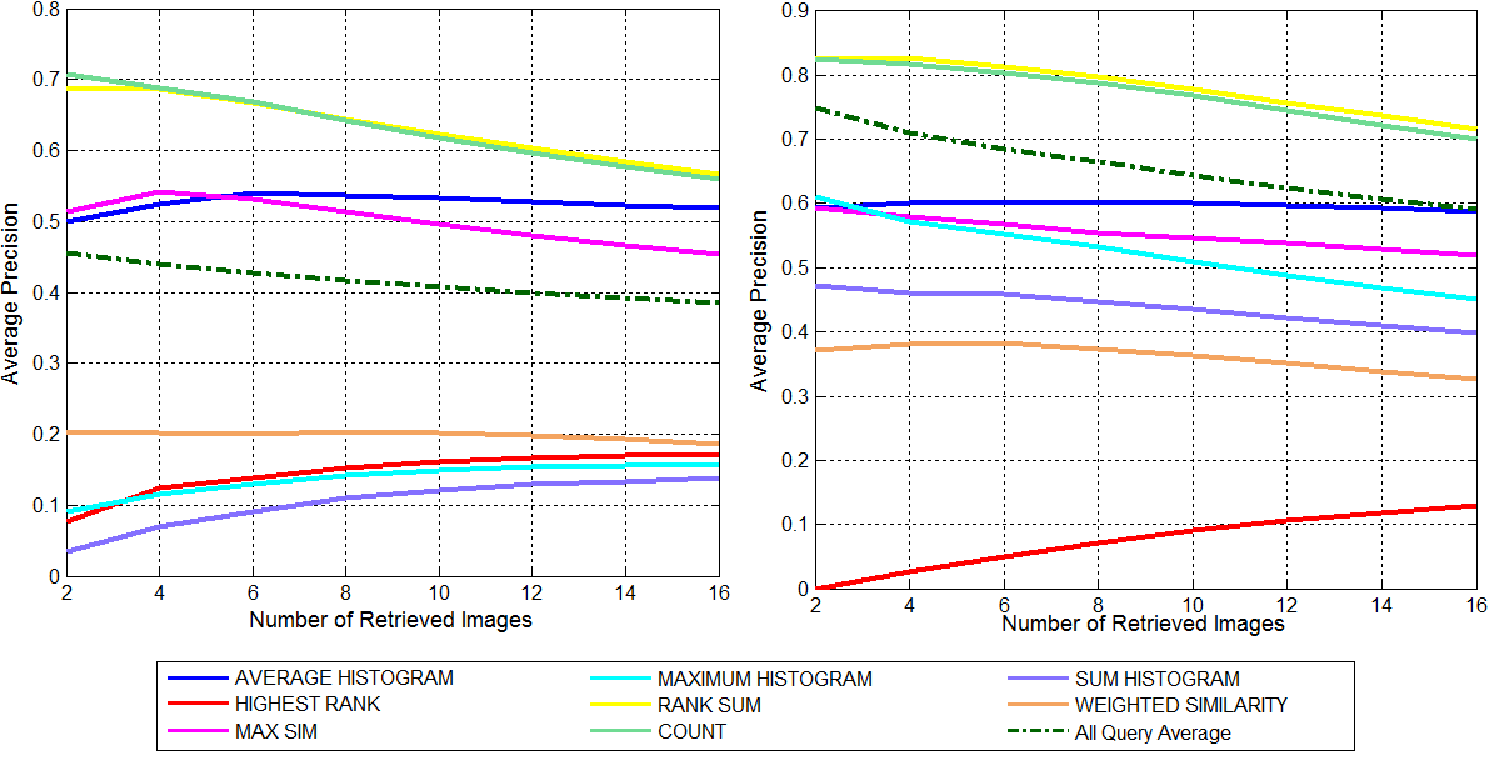}
	\caption{Multi-image query average precision graphs on \textit{Caltech-256} dataset with various early and late fusion methods and the Min-Max Ratio similarity function: (a) background cluttered queries, and (b) clean background queries.}
	\label{fig:AveP-Caltech-multi-image}
\end{figure}

\begin{figure}[h!]
	\centering
	\fbox{\includegraphics[width=1\textwidth]{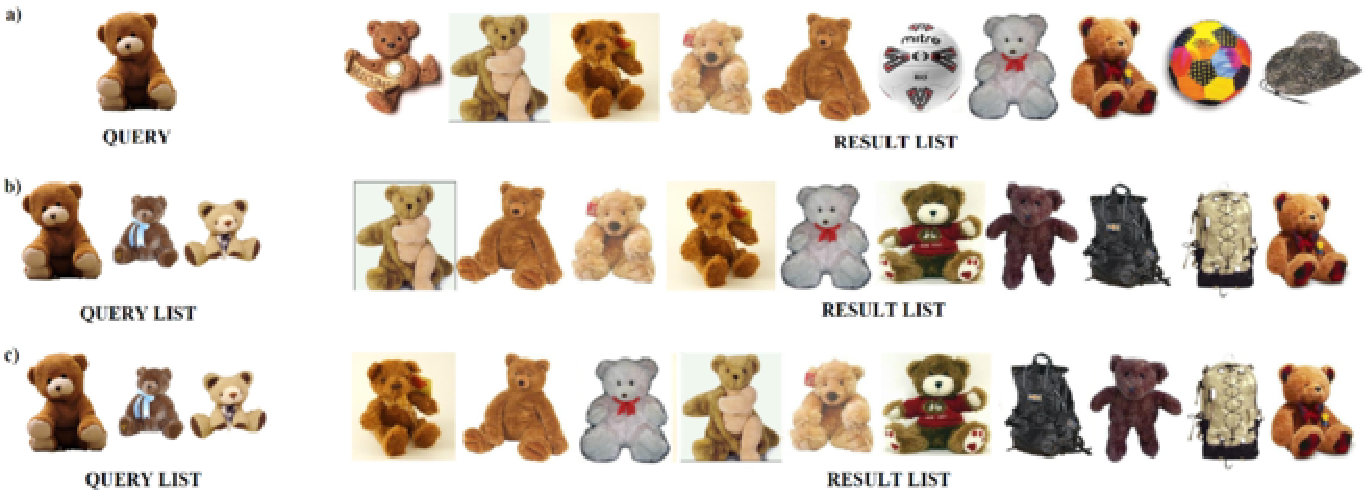}}
	\caption{Single view and multi-image query examples on \textit{Caltech-256} dataset: (a) single view query, (b) multi-image query with \textit{Rank Sum} late fusion method, and (c) multi-image query with \textit{Count} late fusion method. The \textit{Min-Max Ratio} similarity function is used.}
	\label{fig:Caltech256Fig2LateFus}
\end{figure}

\begin{figure}[h!]
	\centering
	\fbox{\includegraphics[width=1\textwidth]{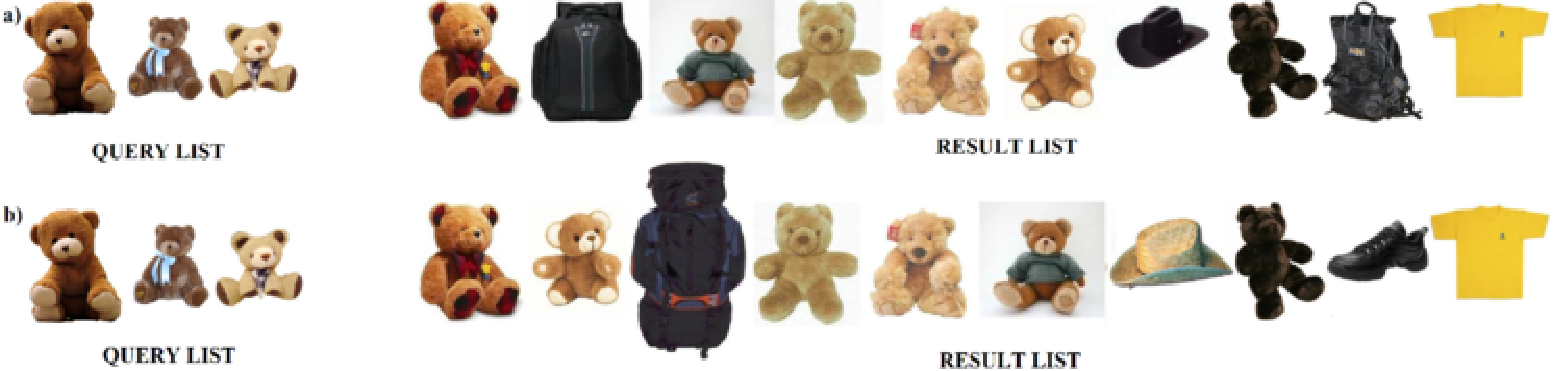}}
	\caption{Multi-image query examples on \textit{Caltech-256} dataset with early fusion: (a) Average Histogram, and (b) Weighted Average Histogram. The \textit{Min-Max Ratio} similarity function is used.}
	\label{fig:Caltech256Fig2EarlyFus}
\end{figure}

Figure~\ref{fig:AveP-Caltech-multi-image} shows the average precision graphs for multi-image queries using various fusion methods and the \textit{Min-Max Ratio} similarity function. The late fusion methods \textit{Rank Sum} and \textit{Count} work better than the other early and late fusion methods. The average precision values are about 0.25 higher on background cluttered queries, and 0.1 higher on clean background queries, compared to the single view queries. Figures~\ref{fig:Caltech256Fig2LateFus} and \ref{fig:Caltech256Fig2EarlyFus} show sample queries.

\begin{figure}[h!]
	\centering
	\includegraphics[width=0.7\textwidth]{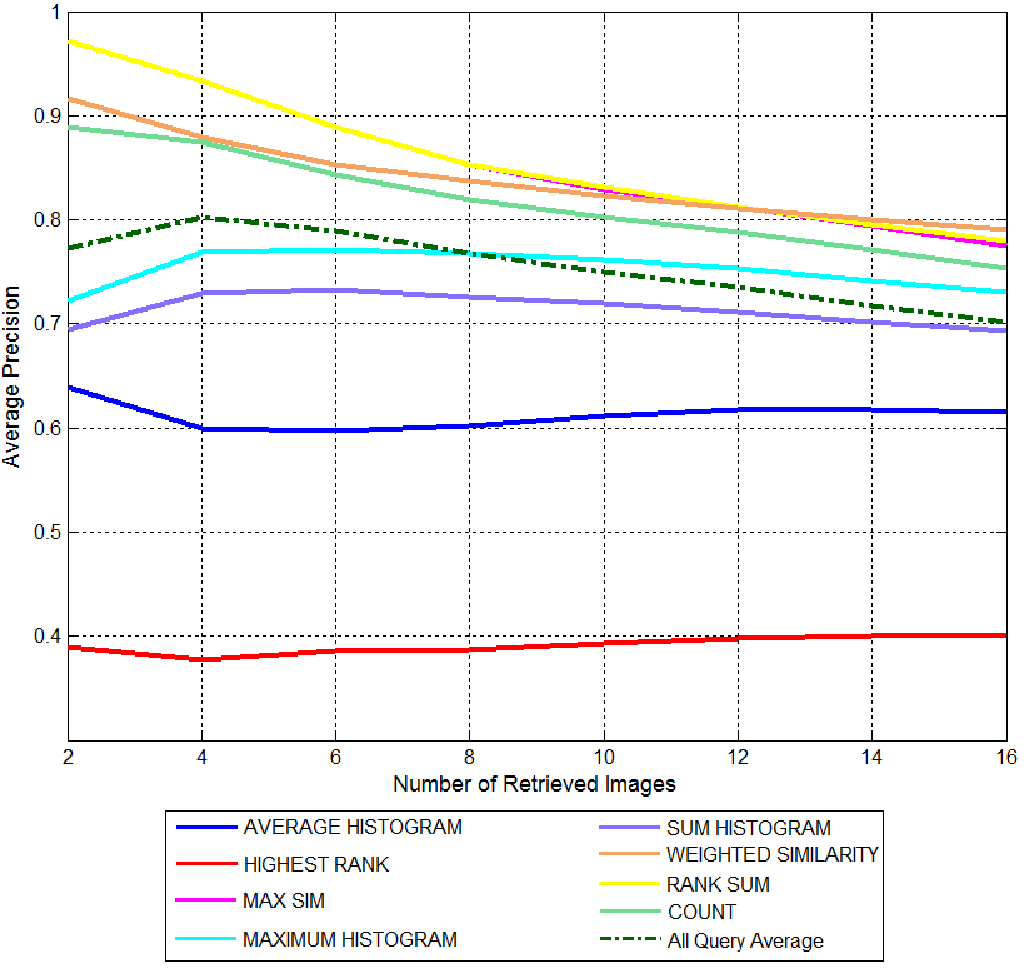}
	\caption{Multi-view query average precision graph on \textit{Caltech-256} dataset with various early and late fusion methods. The \textit{Min-Max Ratio} similarity function is used. The multi-view query images were taken with a mobile phone.}
	\label{fig:AveP-CaltechMobile-mview}
\end{figure}

\begin{figure}[h!]
	\centering
	\fbox{\includegraphics[width=0.7\textwidth]{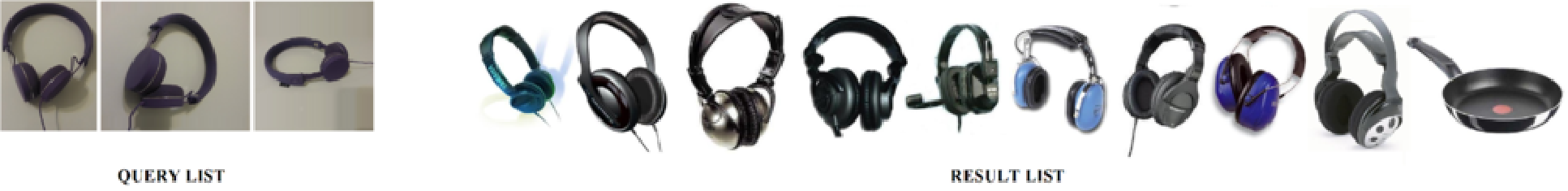}}
	\caption{Multi-view query example on \textit{Caltech-256} dataset with Rank Sum fusion and Min-Max Ratio similarity function. The query images were taken with a mobile phone.}
	\label{fig:Caltech256Fig3Mobile}
\end{figure}

Figures~\ref{fig:AveP-CaltechMobile-mview} and \ref{fig:Caltech256Fig3Mobile} show the average precision graphs and a sample query for multi-view queries using various fusion methods and \textit{Min-Max Ratio} similarity function. As explained above, the multi-view query images of objects were taken with a mobile phone on a clean background. The late fusion methods \textit{Rank Sum}, \textit{Weighted Similarity} and \textit{Count} work better than the other early and late fusion methods. Multi-view queries improve the average precision performance further compared to multi-image queries, since the query images are multiple views of a single object, providing better representation for the query object.

\subsection{Results on the \textit{MVOD} Dataset}
\label{sec:mvod-result}

As mentioned above, \textit{MVOD} is a multi-view dataset we prepared to evaluate the performance of our mobile search system on multi-view object image databases. It is much larger and more challenging than the \textit{Caltech-256} dataset. Since this is a new and completely different dataset, the results are not directly comparable to those of \textit{Caltech-256}.
On this dataset, we performed single view and multi-view query experiments with two types of queries:

\begin{itemize}
 \item \textit{Internet queries.} The multi-view images of $45$ queries, one query per category, are collected from online shopping sites. Similar to the \textit{MVOD} dataset, these query images mostly have clean background.
 \item \textit{Mobile phone queries.} The multi-view query images are obtained with a mobile phone in natural office, home or supermarket environments, in realistic conditions, having adverse effects, like background clutter and illumination problems. The query set has $15$ queries for $15$ categories.
\end{itemize}

The vocabulary size is $10K$ and hard assignment is used for computing the BoW histograms. Single view queries are performed by randomly selecting one of the query views and matching it with one of the database images. Multi-view queries are performed and presented for the best performing similarity function (\textit{Min-Max Ratio}) and best early/late fusion methods based on the above experiments.

\begin{figure}[H]
	\centering
	\includegraphics[width=0.7\textwidth]{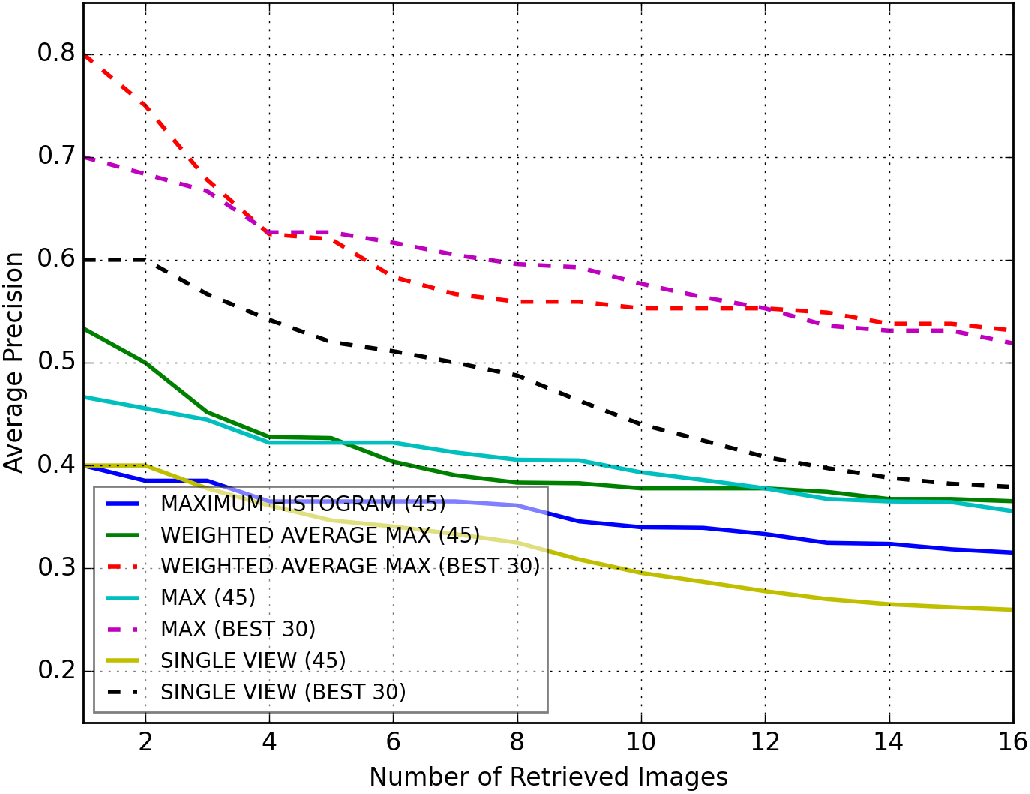}
	\includegraphics[width=0.7\textwidth]{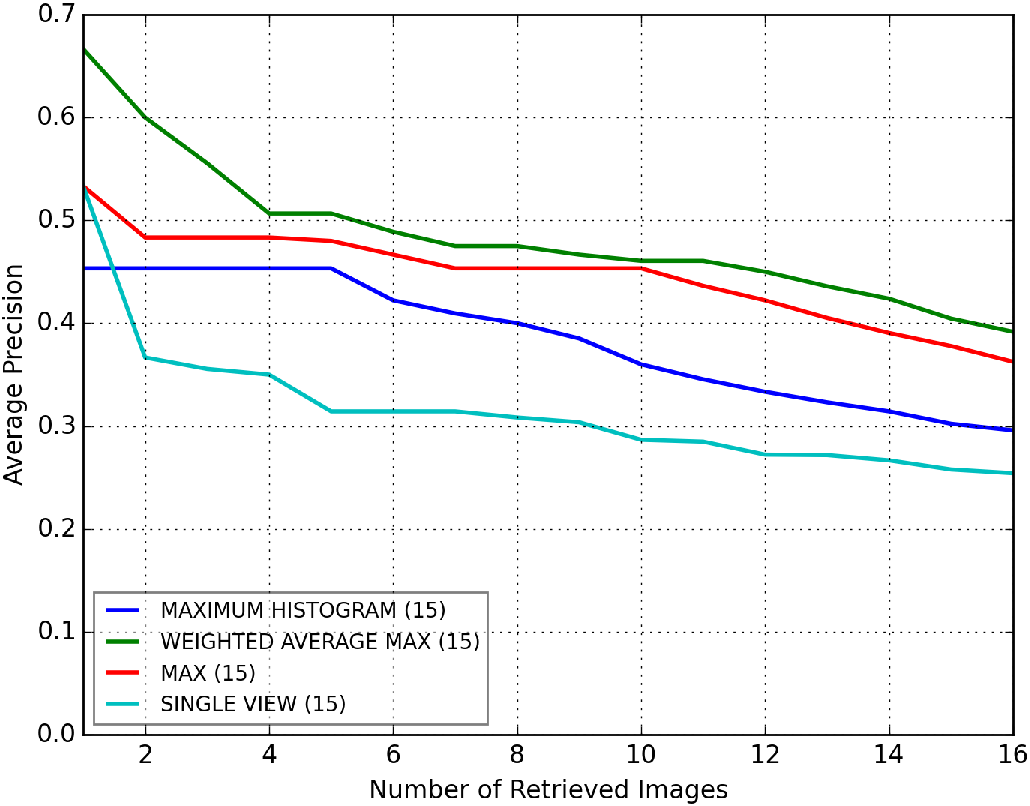}
	\caption{Average precision graphs on the \textit{MVOD} dataset with various early and late fusion methods. Top: query images are from the Internet. Bottom: Query images are taken with a mobile phone. The Min-Max Ratio similarity function is used. The numbers in the legends are the number of queries in the experiment.}
	\label{fig:AveP-Mvod}
\end{figure}

Figure~\ref{fig:AveP-Mvod} shows the average precision graphs for single view and multi-view queries on the \textit{MVOD} dataset. As expected, the average precisions on \textit{MVOD} is lower than those of \textit{Caltech-256}.
Parallel with the \textit{Caltech-256} results, multi-view queries provide an improvement of about $+0.1$ to $+0.2$ over single view queries. The improvement is more on background cluttered queries (taken with a mobile phone), which is important, since, in a real world setting, the query images will usually have background clutter. On the other hand, the average precision for queries with clean background is always higher than queries with cluttered background. It is possible to reduce the influence of background by segmenting out the objects automatically, as in~\cite{34}, or semi-automatically if the user can quickly tap on the screen and select the object of interest, as in~\cite{tapTell}.

\begin{figure}[H]
      \centering
	\fbox{\includegraphics[width=0.93\textwidth]{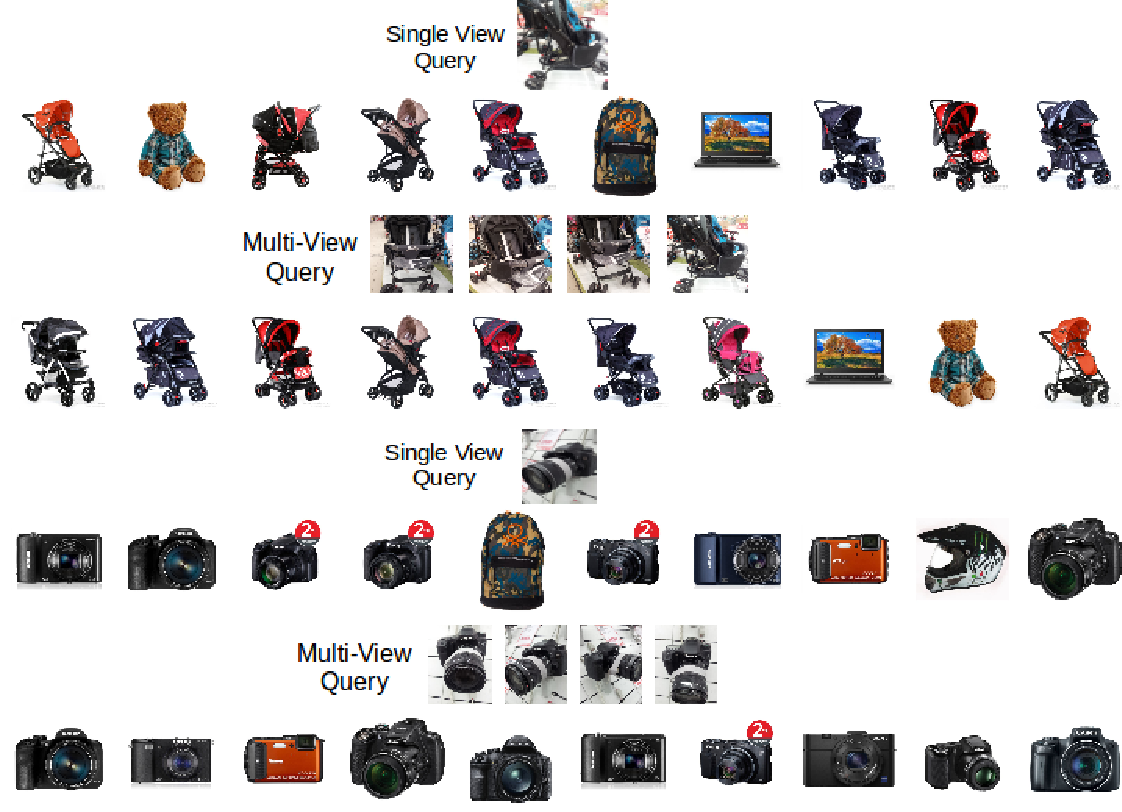}}
	\caption{Single view and multi-view query examples on the \textit{MVOD} dataset, multi-view query with \textit{Max} late fusion method. The query images are taken with a mobile phone.}
	\label{fig:mvod1}
\end{figure}

\begin{figure}[H]
      \centering
	\fbox{\includegraphics[width=0.93\textwidth]{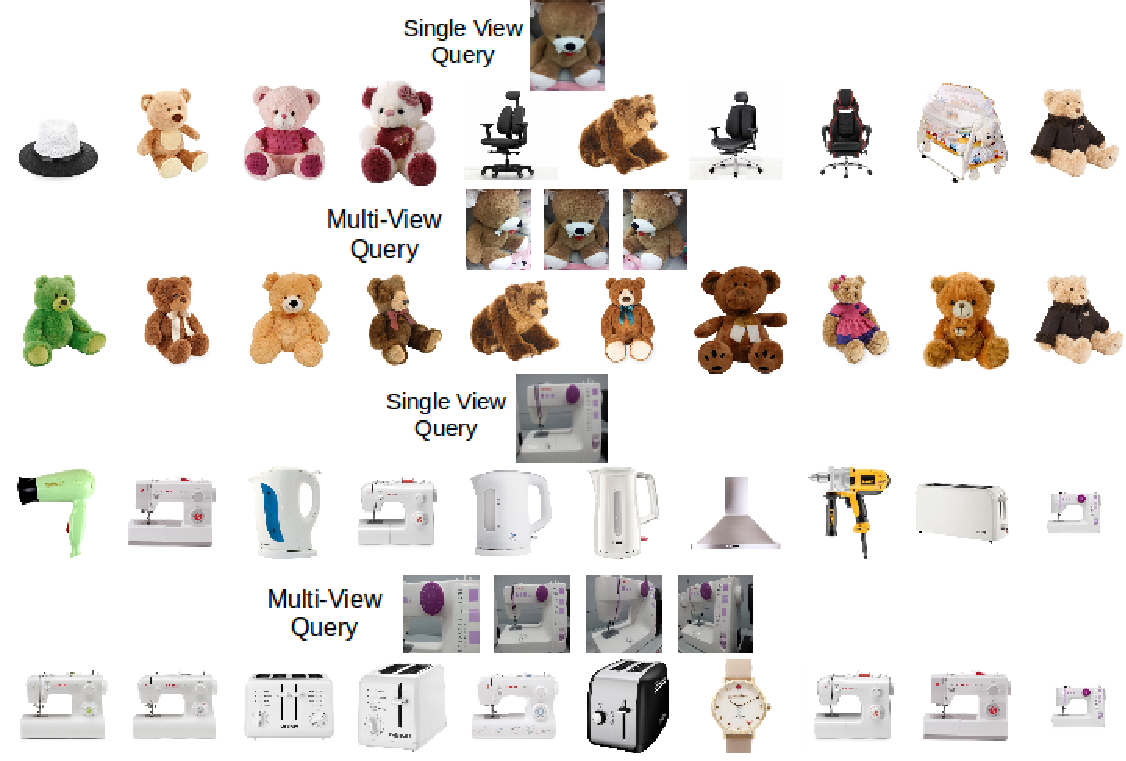}}
	\caption{Single view and multi-view query examples on the \textit{MVOD} dataset, multi-view query with \textit{Weighted Average Max} late fusion method. The query images are taken with a mobile phone.}
	\label{fig:mvod2}
\end{figure}

\begin{figure}[h!]
      \centering
	\fbox{\includegraphics[width=1\textwidth]{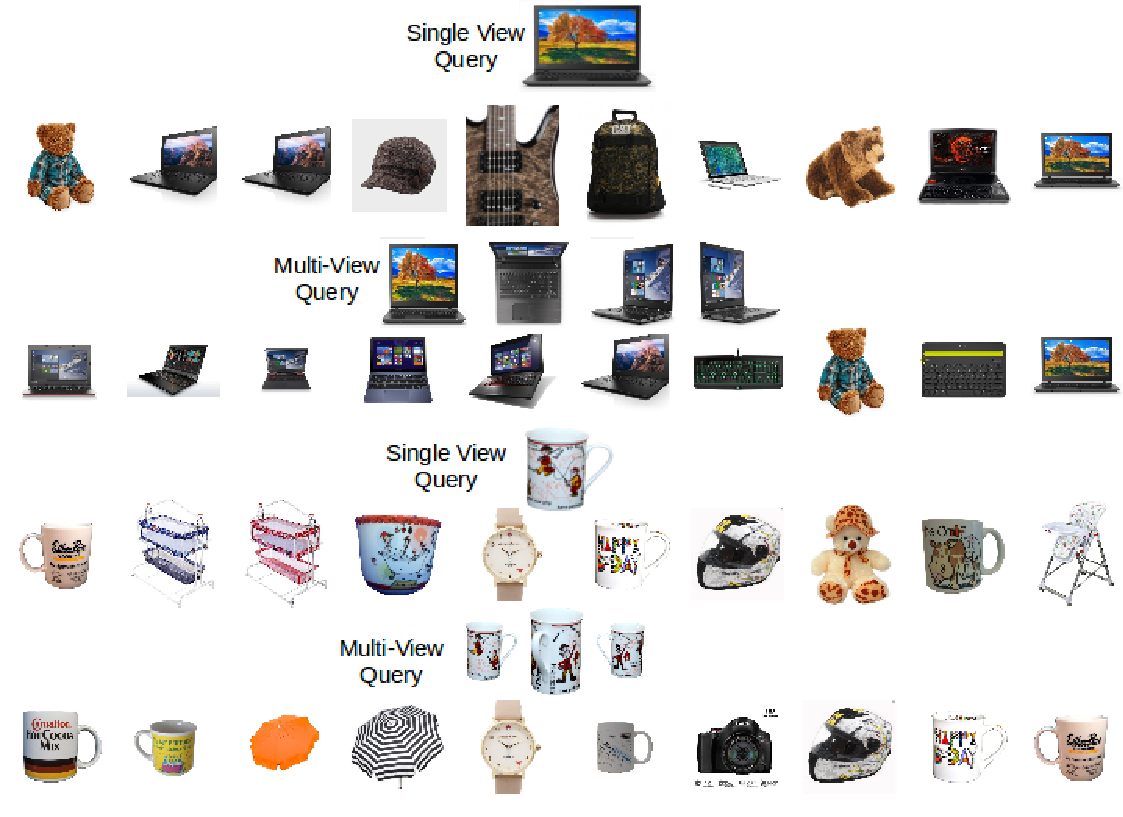}}
	\caption{Single view and multi-view query examples on the \textit{MVOD} dataset, multi-view query with \textit{Max} late fusion method. The query images are from online shopping sites.}
	\label{fig:mvod3}
\end{figure}

Among the fusion methods, the late fusion methods, \textit{Weighted Average of Maximum Similarity} and \textit{Maximum Similarity} work better than others. Sample queries in Figures \ref{fig:mvod1}, \ref{fig:mvod2} and \ref{fig:mvod3} demonstrate the improvement in the result lists for both clean background Internet queries and cluttered background mobile phone queries.

\clearpage

\subsection{Running Time Analysis}
\label{sec:runtime}

Multi-view queries are inherently computationally expensive. In this section, we compare the running times of single and multi-view query methods for different similarity functions. To do so, we measured the time spent for matching the images on the server side; this includes the vector quantization, BoW histogram construction, similarity computation, fusion and ranking. The measurement is done on the \textit{MVOD} dataset, with five queries each having five images, and the measured duration is the average of all queries in each (query) category. Table~\ref{table:timeComparison} summarizes the results. According to the table, the matching times for the similarity functions are close to each other. The increase in running time in multi-view queries is not proportional to the number of images in the query and database, it is lower (due to varying image content and different numbers of interest points detected in each image).
Based on the running times and the average precision performances, the late fusion methods, \textit{Weighted Average of Maximum Similarities} and \textit{Maximum Similarity}, and the early fusion method, \textit{Maximum Histogram}, can be used for multi-view object image search.

\begin{table}
	\centering
	\caption{Running times (ms) of similarity functions and fusion methods.}
	\label{table:timeComparison}
	\resizebox{\textwidth}{!}{
		\begin{tabular}{clccccc}
\hline
			\multicolumn{1}{l}{}  & \multicolumn{1}{c}{}  & \multicolumn{5}{c}{{\bf Similarity Functions}}                                                                    \\
\hline
			\multicolumn{1}{l}{}  &    &
			\begin{tabular}{@{}c@{}}Normalized\\Correlation\end{tabular} &\begin{tabular}{@{}c@{}}Histogram\\Intersection\end{tabular}&\begin{tabular}{@{}c@{}}Normalized Histogram\\Intersection\end{tabular} & Dot Product & Min-Max Ratio \\
\hline
			\multirow{12}{*}{\begin{sideways}\bf{Fusion Methods}\end{sideways}} & \begin{tabular}{@{}c@{}}Single View Query \\ (No Fusion)\end{tabular}
			& 226 & 208  & 242   & 258     & 212  \\
			& Sum Histogram  & 235   & 215   & 322   & 364   & 249  \\
			& Average Histogram & 247  & 261  & 336   & 326  & 258   \\
			& Maximum Histogram & 243  & 294  & 314  & 357     & 253  \\
			& Average Similarity   & 983  & 997  & 1211  & 958  & 994  \\
			& Weighted Avg. Sim. & 996  & 1026  & 1150  & 973  & 1016  \\
			& Maximum Similarity   & 971   & 1015    & 1118 & 964  & 977 \\
			& Average of Max Sim.  & 976  & 1006  & 1125  & 961  & 981  \\
			& Weighted Avg. Max Sim.  & 967   & 1014   & 1147 & 976 & 986
		\end{tabular}
	}
\end{table}

\section{Conclusions and Future Work}
\label{sec:conclusion}

We proposed a new multi-view visual query model on multi-view object image databases for mobile visual search.
We investigated the performance of single view, multi-image and multi-view object image queries on both single view and multi-view object image databases using various similarity functions and early/late fusion methods. We conclude that multiple view images, both in the queries and in the database, significantly improve the retrieval precision. As a result, mobile devices with built-in cameras can leverage the user interaction capability to enable multi-view queries. The performance can be further improved if the query objects are isolated from the background. This can be done automatically as in~\cite{34} or via user interaction, e.g., the user can tap on the screen and select the object-of-interest in the query image~\cite{tapTell}. We implemented a mobile search system and evaluated it on two datasets, both suitable for mobile product search, which is one of the useful application areas of such mobile interactive search systems. Collecting and annotating a large-scale multi-view object image dataset remains as a future work.

Recently, deep convolutional neural networks (ConvNets) have proven to give state-of-the-art results in many computer vision problems, including image classification and retrieval~\cite{dcnn-iccv2015,lcnn-icc2015}. Instead of keypoint-based BoW histograms, ConvNets features can also be used for retrieval in our multi-view object image search framework, since our framework is independent of the features used. 

ConvNets features may be extracted on the mobile device and sent to the server, as in the current architecture, or multi-view images may be sent to the server and all the processing can take place on the server. High-performance ConvNets are usually quite large with millions of parameters~\cite{inception-cnn-2015} and require a high amount of processing power and memory. This is a serious limitation for a mobile search system; using large networks on current mobile devices is not feasible due to to the stringent memory limits on the running processes. Smaller networks, on the other hand, may not give satisfactory performance. The second alternative, sending images to the server, may require a large amount of uplink data traffic, which may be both costly and slow (upload data rates are much lower than download data rates). In summary, an interesting research direction is to design a mobile search system architecture that can use the state-of-the-art ConvNets efficiently.

\begin{acknowledgements}
The first author was supported by The Scientific and Technological Research Council of Turkey (T\"{U}B\.{I}TAK) under B\.{I}DEB 2228-A Graduate Scholarship.
\end{acknowledgements}

\bibliographystyle{spmpsci}      
\bibliography{References}

\end{document}